%% file: LSSG-general.tex
\documentclass[11pt]{article}
\usepackage{fullpage}

\usepackage{microtype}
\usepackage{graphicx,amsfonts,amsmath,amssymb,epsfig,hyperref,color,paralist}
\usepackage{algorithm}

\title{A Centralized Local Algorithm for the Sparse Spanning Graph Problem}
\author{
		Christoph Lenzen
		\footnote{MPI for Informatics, Saarbr\"{u}cken 66123, Germany, \textit{clenzen@mpi-inf.mpg.de}}
		\and  Reut Levi
		\footnote{ MPI for Informatics, Saarbr\"{u}cken 66123, Germany, \textit{rlevi@mpi-inf.mpg.de}}
	}

\newcommand{\eps}{\epsilon}
\newcommand{\eqdef}{\stackrel{\rm def}{=}}
\newtheorem{defn}{Definition}         
\newcommand{\BD}{\begin{defn}} \newcommand{\ED}{\end{defn}}
\newcommand{\BE}{\begin{enumerate}} \newcommand{\EE}{\end{enumerate}}
\newcommand{\BI}{\begin{itemize}} \newcommand{\EI}{\end{itemize}}
\newcommand{\calA}{{\cal A}}
\newcommand{\calC}{{\cal C}}

\newtheorem{thm}{Theorem}
\newcommand{\BT}{\begin{thm}} \newcommand{\ET}{\end{thm}}
\def\FullBox{\hbox{\vrule width 8pt height 8pt depth 0pt}}
\newcommand{\ourqed}{\;\;\;\FullBox}
\newenvironment{ourproof}{\noindent{\bf Proof:~~}}{\(\ourqed\)}
\newcommand{\BPF}{\begin{ourproof}} \newcommand {\EPF}{\end{ourproof}}
\newenvironment{proofof}[1]{\noindent{\bf Proof of {#1}:~~}}{\(\ourqed\)}
\newcommand{\BPFOF}{\smallskip \begin{proofof}} \newcommand {\EPFOF}{\end{proofof}}
\newcommand{\BEQ}{\begin{equation}} \newcommand{\EEQ}{\end{equation}}
\newcommand{\BEQN}{\begin{eqnarray}}\newcommand{\EEQN}{\end{eqnarray}}
\newtheorem{lem}{Lemma}      
\newcommand{\BL}{\begin{lem}} \newcommand{\EL}{\end{lem}}
\newtheorem{cor}[thm]{Corollary}      
\newcommand{\BC}{\begin{cor}} \newcommand{\EC}{\end{cor}}
\newtheorem{clm}[lem]{Claim}
\newcommand{\BCM}{\begin{clm}} \newcommand{\ECM}{\end{clm}}
\newcommand{\poly}{{\rm poly}}
\newtheorem{fact}{Fact}      
\newcommand{\BF}{\begin{fact}} \newcommand{\EF}{\end{fact}}

\newcommand{\cent}{\sigma}
\newcommand{\W}{W}  
\newcommand{\cell}{C}  
\newcommand{\chiW}{\chi_{\mbox{\tiny\it W}}}

\newcommand{\Vor}{{\rm Vor}}
\renewcommand{\Pr}{\mathrm{Pr}}
\newcommand{\Exp}{\mathrm{Exp}}

\newcommand{\s}[1]{\left\lvert #1 \right\rvert}

\renewcommand{\d}{d} 
\newcommand{\dnew}[1]{{#1}}

\newcommand{\calT}{{\cal T}}

\begin{document}

\maketitle

\begin{abstract}
\input{abstract}
\end{abstract}

\input{intro}
\section{Preliminaries}\label{sec:prel}
The graphs we consider are undirected and have a known degree bound $\Delta$,
and we assume we have query access to their incidence list representation.
Namely, for any vertex $v$ and index $1 \leq i \leq \Delta$, it is possible to obtain
the $i^{\rm th}$ neighbor of $v$ by performing a query to the graph
(if $v$ has less than $i$ neighbors, then
a special symbol is returned).
Without loss of generality, we assume that graphs are simple, i.e., contain neither loops nor parallel edges.\footnote{The answer on a self-loop can always be negative,
and we can default to rejecting all but the first edge between two nodes.}
The number of vertices in the graph is $n$ and we assume that each vertex $v$ has a unique id, which for simplicity we also denote by $v$.
There is a total order on the ids, i.e., given any two distinct ids $u$ and $v$, we can decide whether $u<v$ or $v<u$.

Let $G = (V,E)$ be a graph, where $V = [n]$.
We denote the distance between two vertices $u$ and $v$ in $G$ by $d_G(u,v)$.
For vertex $v \in V$ and an integer $r$,
let $\Gamma_r(v,G)$ denote the set of vertices at distance at most
$r$ from $v$.
When the graph $G$ is clear from the context, we shall use the shorthands
$d(u,v)$ and $\Gamma_r(v)$ for $d_G(u,v)$ and $\Gamma_r(v,G)$, respectively.

The total order on the vertices induces a total order $r$ on
the edges of the graph in the following straightforward manner:
$r(\{u,v\}) < r(\{u',v'\})$ if and only if $\min\{u,v\} < \min\{u',v'\}$ or $\min\{u,v\} = \min\{u',v'\}$
and $\max\{u,v\} < \max\{u',v'\}$.
The total order over the vertices also induces an order over those vertices visited by
a Breadth First Search (BFS) starting from any given vertex $v$, and whenever we refer to
a BFS, we mean that it is performed according to this order.

Whenever referring to one of the above orders, we may refer to the \emph{rank} of an element in the respective order.
This is simply the index of the respective element when listing all elements ascendingly with respect to the order.

For a graph $G = (V,E)$ and a pair of disjoint subsets of vertices $A\subset V$ and $B\subset V$ let $E_G(A, B) \eqdef \{(u,v)\in E\,|\, u\in A \wedge v\in B\}$. When it is clear from the context, we omit the subscript.
We say that a pair of subsets of vertices $A$ and $B$ are {\em adjacent} if $E_G(A, B) \neq \emptyset$.

\section{An Algorithm that Works under a Promise}\label{sec:promise}
We begin by describing an LSSG algorithm which works under the following promise on the input graph $G = (V, E)$.
Sample $\ell$ uniformly at random from $ [2\log n/ \log(1+ \eps),2\log n/ \log(1+ \eps)+\Delta/\eps\}]$, and let $k \eqdef cn^{1/3} \ln n \cdot \ell\Delta/\eps$, where $c$ is a constant that will be determined later.
For every $v \in V$, let $i_v \eqdef \min_r \{|\Gamma_r(v)| \geq k\}$. 
We are promised that $\max_{v\in V} \{i_v\} \leq \ell$. In words, we assume that the $\ell$-hop neighborhood of every vertex in $G$ contains at least $k$ vertices. 
First, we fix a simple partition of $V$.

\begin{figure}[t!]
\centering
\includegraphics[width=.9\textwidth]{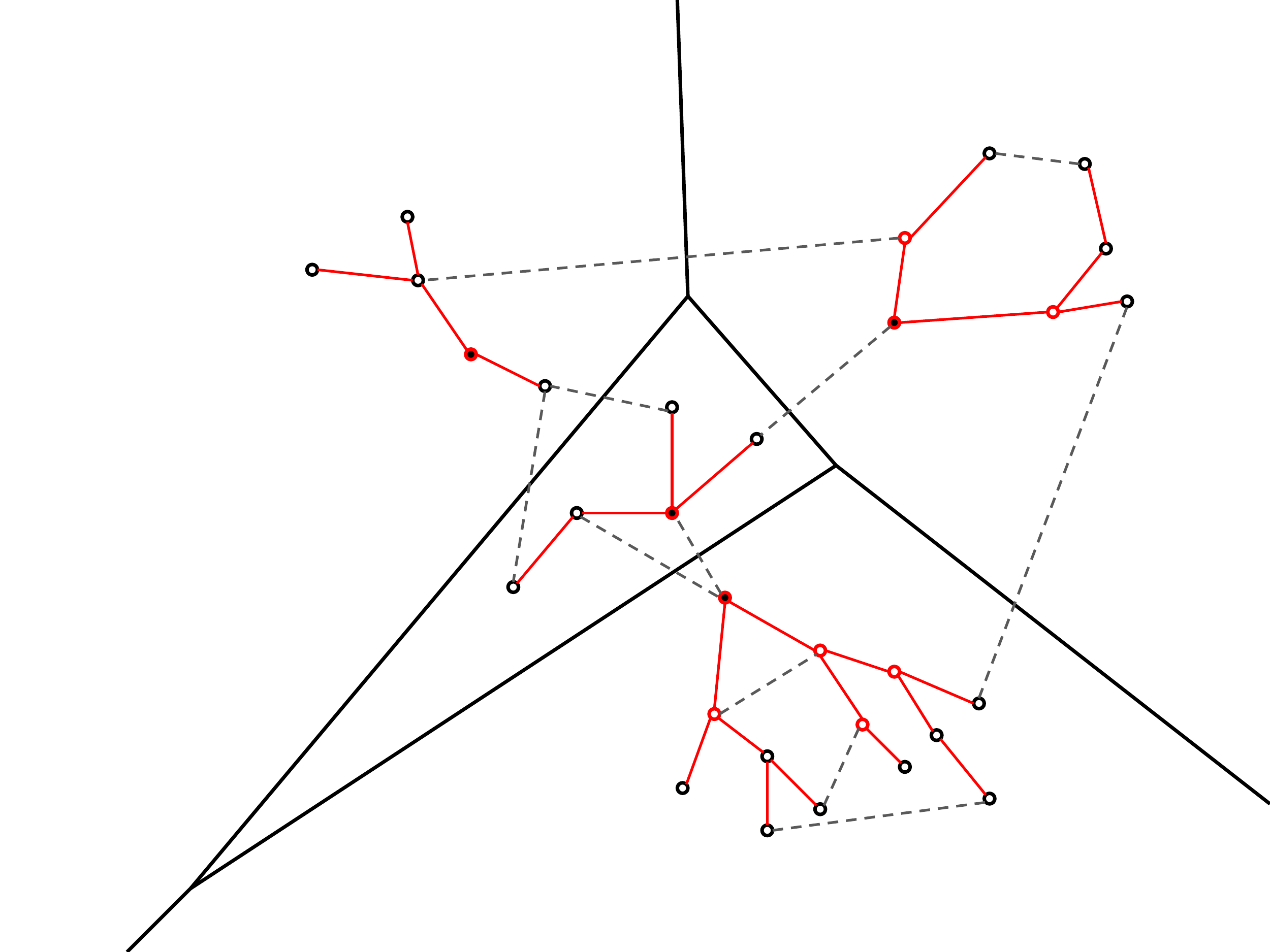}
\caption{Partition of a graph into cells and clusters, where $k=6$. Black lines are the borders of Voronoi cells, whose centers have black fillings. Red edges belong to the BFS trees spanning the clusters, while dashed gray lines are non-tree edges. Red circles indicate singleton clusters (if the node has a red child) or the roots of subtrees that form a cluster (if the children are black).}\label{fig:cells_clusters}
\end{figure}

\subsection{The Underlying Partition}\label{sub.par}

\subparagraph*{Centers.}
Pick a set $S\subset |V|$ of $r \eqdef \Theta(\eps n^{2/3}/\ln n)$ vertices at random. 
We shall refer to the vertices in $S$ as {\em centers}.  
For each vertex $v \in V$, its {\em center}, denoted by $c(v)$, is the center which is closest to $v$ amongst all centers (break ties between centers according to the id of the center).

\subparagraph*{Voronoi cells.}
The {\em Voronoi cell} of a vertex $v$, denoted by $\Vor(v)$, is the set of all vertices $u$ for which $c(u) = c(v)$.
Additionally, we assign to each cell a random rank, so that there is a uniformly random total order on the cells;
note carefully that the rank of a cell thus differs from the rank of its center (which is given by its identifier, which is not assigned randomly).
We remark that we can determine the rank of the cell from the shared randomness and the cell's identifier, for which we simply use the identifier of its center. 

\subparagraph*{Clusters.}
For each Voronoi cell, consider the BFS tree spanning it, which is rooted at the respective center. 
For every $v\in V$, let $p(v)$ denote the {\em parent} of $v$ in this BFS tree. If $v$ is a center then $p(v) = v$.
For every $v\in V\setminus S$, let $T(v)$ denote the subtree of $v$ in the above-mentioned BFS tree when we remove the edge $\{v, p(v)\}$;
for $v\in S$, $T(v)$ is simply the entire tree.
Now consider a Voronoi cell.
If the cell contains at most $k$ vertices, then the {\em cluster} of all the vertices in the Voronoi cell is the cell itself.
Otherwise, there are two cases. 
If $T(v)$ contains at least $k$ vertices, then the cluster of $v$ is the singleton $\{v\}$.
Otherwise, $v$ has a unique ancestor $u$ (including $v$) for which $T(u)$ contains less than $k$ vertices and $T(p(u))$ contains at least $k$ vertices. 
The cluster of $v$ is the set of vertices in $T(u)$. 
For a cluster $C$, let $c(C)$ denote the center of the vertices in $C$ (all the vertices in the cluster have the same center).
Let $\Vor(C)$ denote the Voronoi cell of the vertices in $C$.

\medskip\noindent 
This describes a partition of $V$ into Voronoi cells, and a refinement of this partition into clusters.
See Figure~\ref{fig:cells_clusters} for an illustration.

\subsection{The Edge Set}\label{subsec.edge}
Our spanner, $H = (V, E')$, initially contains, for each Voronoi cell $\Vor$ the edges of the BFS tree that spans $\Vor$, i.e., the BFS tree rooted at the center of $\Vor$ spanning the subgraph induced by $\Vor$ (see Section~\ref{sec:prel} for more details). 
Clearly, these edges also span the clusters. 
Next, we describe which edges we add to $E'$ in order to connect the clusters.
\subsubsection*{Marked Clusters and Clusters-of-Clusters}
Each center is {\em marked} independently with probability $p \eqdef 1/n^{1/3}$. 
If a center is marked, then we say that its Voronoi cell is marked and all the clusters in this cell are marked as well.

\begin{figure}[t!]
\centering
\includegraphics[width=.7\textwidth]{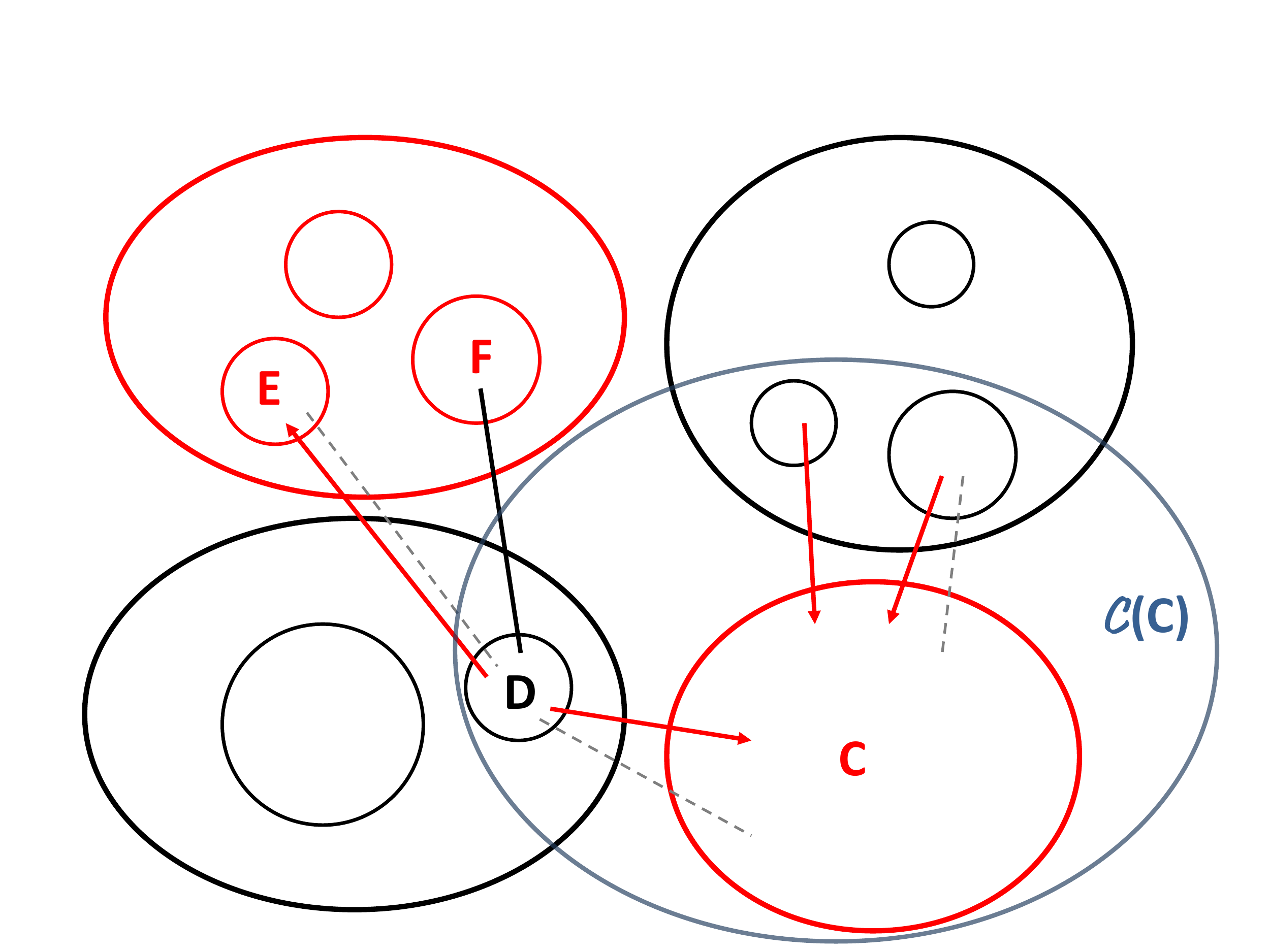}
\caption{Illustration of marked clusters and clusters of clusters. Thick red and black ovals are marked and unmarked cells, respectively. Thin circles are clusters, where cluster $C$ comprises its entire cell. Thick edges are the ones of minimum rank between their incident clusters, while the dotted edges do not meet this criterion. The arrows of red edges indicate participation in the respective adjacent marked cluster; note that $D$ does not participate in $\calC(F)$, as for each adjacent marked cell $\Vor$ it exclusively participates in the cluster-of-clusters connected to it by the edge of minimum rank in $E(D,\Vor)$. $\calC(C)$ is marked in blue; all its constituent clusters also participate in $\calC(C)$, as $\Vor(C)=C$.}\label{fig:clustersofclusters}
\end{figure}

\subparagraph*{Cluster-of-clusters.}
For every marked cluster, $C$, define the {\em cluster-of-clusters} of $C$, denoted by $\calC(C)$, to be the set of clusters which consists of $C$ and all the clusters which are adjacent to $C$. 
A cluster $B$ is {\em participating} in $\calC(C)$ if $B \in \calC(C)$ and the edge of minimum rank in $E(B, \Vor(C))$ also belongs to $E(B, C)$.
Thus, if $B$ is adjacent to $\Vor(C)$ and $\Vor(C)$ is marked, then there is a unique cluster $D\subseteq \Vor(C)$ such that $B$ participates in $\calC(D)$.
See Figure~\ref{fig:clustersofclusters} for a visualization.

\subsubsection*{The Edges between Clusters}
By saying that we {\em connect} two adjacent subsets of vertices $A$ and $B$, we mean that we add the minimum ranked edge in $E(A, B)$ to $E'$.
For a cluster $A$, define its {\em adjacent centers} $\Vor(\partial A)\eqdef \{\Vor(v)\,|\,u\in A \wedge \{u,v\}\in E \}\setminus \{\Vor(A)\}$, i.e., the set of Voronoi cells that are adjacent to $A$.
This definition explicitly excludes $\Vor(A)$, as there is no need to connect $A$ to its own Voronoi cell.

\begin{figure}[t!]
\centering
\includegraphics[width=.7\textwidth]{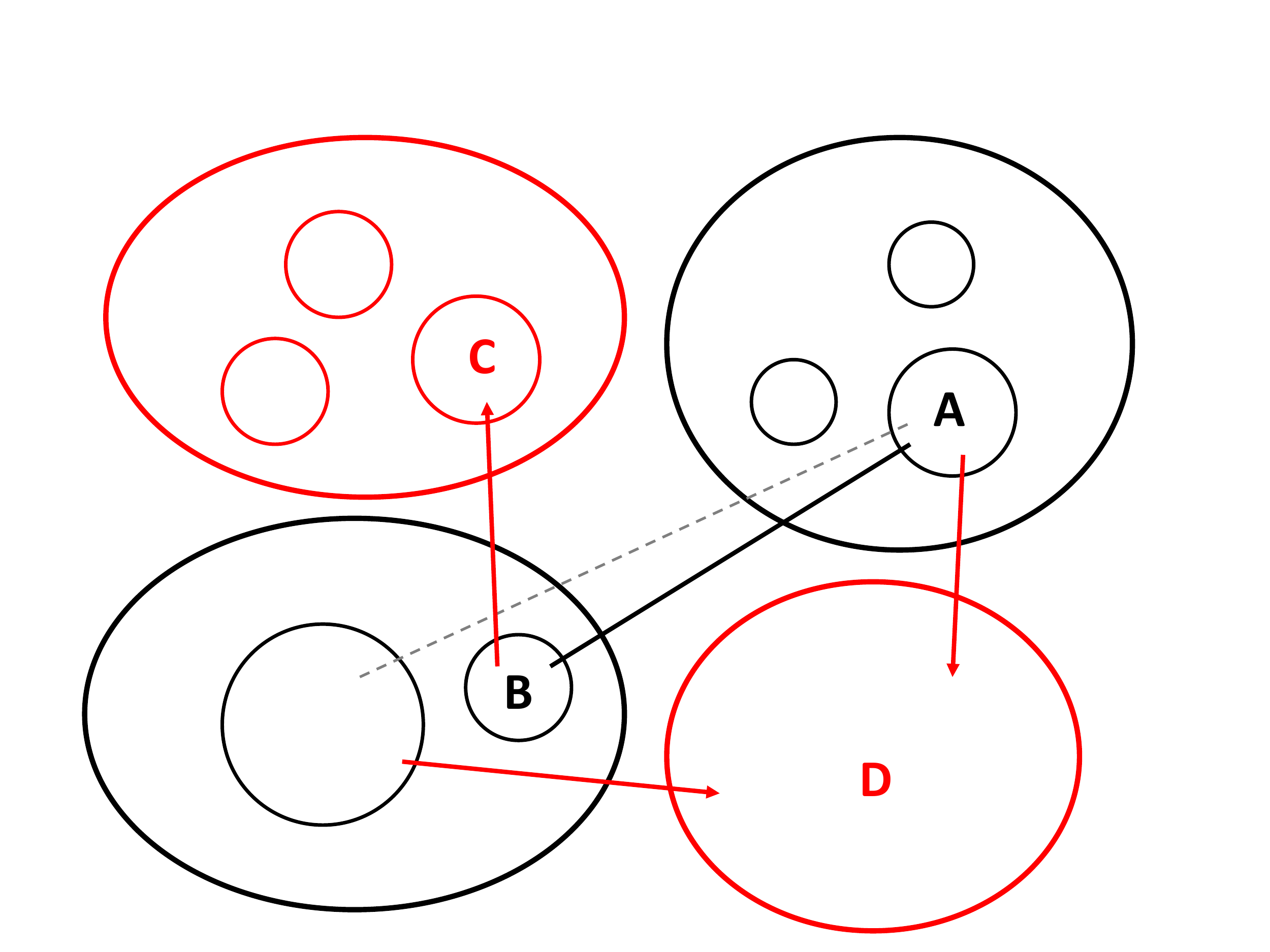}
\caption{Illustration of the third edge selection rule. In this example, the thick black edge has minimum rank in $E(A,\Vor(B))$ and $B$ has minimum rank in $\partial A \cap \partial C$. However, the rank of $\Vor(D)$ is smaller than that of $\Vor(B)$, and hence the dashed edge is not selected. Here, $A$ will select the direct edge connecting it to $D$, due to the first rule.}\label{fig:rule3}
\end{figure}

We next describe how we connect the clusters. The high-level idea is to make sure that every marked cluster and the clusters that participate in the respective cluster-of-clusters remain connected. For clusters which are not adjacent to any marked cluster we make sure to keep them connected to all adjacent Voronoi cells. Formally:

\begin{compactenum}
\item We connect every cluster to every adjacent marked cluster.\label{connect_marked}
\item Each cluster $A$ that is not participating in any cluster-of-clusters (i.e., no cell adjacent to $A$ is marked) we connect to each adjacent cell.
\label{connect_no_adjacent}
\item Suppose cluster $A$ is adjacent to cluster $B$, where $B$ is adjacent to a marked cell $\Vor$.
Denote by $C$ the (unique) cluster in cell $\Vor$ for which $B$ participates in $\calC(C)$.
We connect $A$ with $B$ if the following conditions hold:
\begin{compactitem}
\item $\Vor(B)$ has minimum rank amongst $\Vor(\partial A) \cap \Vor(\partial C)$
\item the minimum ranked edge in $E(A, \Vor(B))$ is also in $E(A, B)$
\end{compactitem}
\label{connect_indirect}
\end{compactenum}
Figure~\ref{fig:rule3} showcases the third rule.

\subsection{Sparsity} 
\BL
The number of clusters, denoted by $s$, is at most $|S| + n \ell (\Delta + 1)/k$.
\EL
\BPF
We first observe that, due to the promise on $G$, it follows that for every $v \in V$, $\d(v, c(v)) \leq \ell$ w.h.p.
Recall the terminology from Subsection~\ref{sub.par}. 
Consider $v$ for which $|T(v)| \geq k$ and therefore its cluster is the singleton $\{v\}$.  
We say that a vertex $u$ is {\em special} if $|T(u)| \geq k$ and for every child of $u$ in $T(u)$, $t$, it holds that $|T(t)| < k$.
By an inductive argument, it follows that $v$ is an ancestor of a special vertex.
Since for every pair of special vertices $u$ and $w$, $T(u)$ and $T(w)$ are vertex disjoint, we obtain that there are at most $n/k$ special vertices. Since for every special vertex, there are at most $\ell$ ancestors, the total number of vertices $v$ with $|T(v)|\geq k$ is bounded by $n\ell / k$.

Observe that any cluster either (i) is a singleton $\{v\}$ with $|T(v)|\geq k$, (ii) contains a node $v$ such that $|T(p(v))|\geq k$, or (iii) is an entire Voronoi cell.
We just bounded the number of clusters of type (i) by $n\ell /k$, and immediately get a bound on the number of type (ii) clusters of $n\ell \Delta / k$.
The number of type (iii) clusters is bounded by the number of Voronoi cells $|S|$, showing the desired bound on $s$.
\EPF
\BL
$\Exp(|E'|) \leq (1+O(\eps))|V|$.
\EL
\BPF
The number of edges we add due to the BFS trees of the Voronoi cells is at most $|V|-1$.

The number of edges which are taken due to Condition~\ref{connect_marked} is at most $s$ times the number of marked clusters.
In expectation, there are $s\cdot p$ marked clusters, yielding at most $s^2 p$ edges in expectation.
Since $s = O(\eps n^{2/3}/\ln n)$ and $p = 1/n^{1/3}$ we obtain that $s^2 p = O(\eps n /\ln n)$.

Let $A$ be a cluster. The number of edges which are adjacent to $A$ and are taken due to Condition~\ref{connect_indirect} is bounded by the total number of clusters-of-clusters. The number of clusters-of-clusters is exactly the number of marked clusters.
Thus, the total number of edges which are taken due to Condition~\ref{connect_indirect} is bounded by $s^2 p$.

Observe that the probability that cluster $A$ is not adjacent to a marked cell is $(1-p)^{|\Vor(\partial A)|}\leq e^{-p|\Vor(\partial A)|}$.
Hence, if $|\Vor(\partial A)|\geq 3p^{-1}\ln n$, $A$ is adjacent to a marked cell w.h.p.
Using a union bound over all clusters, it follows that w.h.p.\ each cluster $A$ without an adjacent marked cell satisfies that $|\Vor(\partial A)|\leq 3p^{-1}\ln n$ with probabilty at least $1-1/n^2$;
the probability at most $1/n^2$ event that this bound is violated cannot contribute more than $|E|/n^2<1$ to the expectation.
Therefore, the total number of edges which are taken due to Condition~\ref{connect_no_adjacent} is bounded by $(3s\ln n)/p + 1 = O(\eps n)$.

To conclude, the total number of edges in $E'$ is at most $|V| (1+ O(\eps))$ in expectation, as desired.
\EPF

\input{conn}
\input{alg-general}

\section{The Local Algorithm}\label{sec:localC}
In this section we prove the our main theorem.

\begin{algorithm}[t]
\caption{LSSG for general graphs}\label{alg:main}
\textbf{Input:} $\{u, v\} \in E$\\
\textbf{Output:} whether $\{u, v\}$ is in $E'$ or not.
\BE 
\item If $u, v \in R$, compute the output of algorithm of Elkin and Neiman at $u$ and $v$ when running it on the connected component of $u$ and $v$ in the subgraph induced by $R$. 
Return {\bf true} if $\{u, v\} \in C(u)\cup C(v)$ and {\bf false} otherwise.\label{s1}
\item If $\{u, v\}  \in E(R,\bar{R})$, return {\bf true}.\label{s2}
\item Otherwise, $u,v\in \bar{R}$ and we proceed according to Section~\ref{sec:promise}, where all nodes in $R$ are ignored:
\BE
\item If $\Vor(u)  = \Vor(v)$, return {\bf true} if $\{u, v\}$ is in the BFS tree of $\Vor(u)$ and {\bf false} otherwise.  
\item Otherwise, let $Q$ and $W$ denote the clusters of $u$ and $v$, respectively.
Return {\bf true} if at least one of the following conditions hold for $A = Q$ and $B = W$, or symmetrically, for $A = W$ and $B = Q$, and {\bf false} otherwise. 
\begin{enumerate}
\item \label{con.0} $A$ is a marked cluster and $\{u,v\}$ has minimum rank amongst the edges in $E(A,B)$.
\item \label{con.1} $A$ is not participating in any cluster-of-clusters. Namely, all the clusters which are adjacent to $A$  are not marked. 
In this case, we take $\{u, v\}$ if it has minimum rank amongst the edges in $E(A, \Vor(B))$.
\item \label{con.2} There exists a marked cluster $C$ such that $A$ is participating in $\calC(C)$, and the following holds:
\begin{itemize}
\item $\Vor(A)$ has minimum rank amongst $\Vor(\partial B) \cap \Vor(\partial C)$
\item $\{u,v\}$ has minimum rank amongst the edges in $E(B, \Vor(A))$.  
\end{itemize}
\end{enumerate}
\EE
\EE 
\end{algorithm}
\BT\label{thm:alg}
Algorithm~\ref{alg:main} is an LSSG algorithm. 
For any graph $G$ over $n$ vertices of maximum degree at most $\Delta$ and $\eps > 0$,
its query complexity, space complexity (length of the random seed), and running time are $\tilde{O}(n^{2/3}) \cdot \poly(1/\eps, \Delta)$.
\ET
\BPF
The correctness of the algorithm follows from the previous sections. We shall prove that its complexity is as claimed.
We analyze the complexity in terms of $n$. There are additional factors that depend polynomially in $\Delta$ and $1/\eps$.
The following claims hold w.h.p., simultaneously, for all vertices.

Recall that a vertex $u$ is remote if $\Gamma_\ell(u)$ does not contain a center.
Due to the sampling probability for centers, a center is found after exploring $\tilde{O}(n^{1/3})$ vertices.
Therefore, we can decide for any vertex $u$ whether it is in $R$ with query and time complexity $\tilde{O}(n^{1/3})$.
Moreover, if $u\in \bar{R}$, without additional cost the respective subroutine can return $c(u)$, the center of $u$, and $\d(u,c(u))$ (as we explore in a BFS fashion).

For Step~\ref{s1}, we need to determine $\Gamma_\ell(x)\cap R$.
Since $|\Gamma_\ell(u)| = \tilde{O}(n^{1/3})$ for every vertex $u\in R$, 
we obtain that the query and time complexity of this step is $\tilde{O}(n^{2/3})$, in total.
Accordingly, Step~\ref{s2} has query and time complexity $\tilde{O}(n^{1/3})$.

If $u,v\in \bar{R}$, the algorithm proceeds as in Section~\ref{sec:promise}.
We claim that reconstructing the clusters of $u$ and $v$, determining the centers of all adjacent nodes, and deciding whether $\{u,v\}$ is a BFS edge (i.e., $\Vor(u)=\Vor(v)$ and $u$ is the parent of $v$ in the BFS of $\Vor(u)$ rooted at $c(u)$ or vice versa) takes $\tilde{O}(n^{2/3})$ queries and time.

We show first that we can reconstruct clusters efficiently.
W.l.o.g., consider $u$.
We determine $c(u)$ and $\d(u,c(u))$.
For each neighbor $w$ of $u$, we determine whether it is in $\bar{R}$ (if not, the node is discarded), its center $c(w)$, and the distance $\d(w,c(w))$.
As $u\in \bar{R}$, and assuming that $u \neq c(u)$, it must have at least one neighbor $w$ in distance $\d(u,c(u))-1$ of $c(u)$;
any such $w$ satisfies that $c(w)=c(u)$, as otherwise $d(u,c(w))=d(w,c(w))$ and $c(w)<c(u)$, a contradiction to the tie-breaking rule for centers.
Among these candidates $w$, we know that the one with minimum rank is the parent of $u$ in the BFS tree of $\Vor(u)$ rooted at $c(u)$, due to the tie-breaking rule for the BFS construction.
Otherwise, we can use this subroutine to partially explore the BFS of $\Vor(u)$:
given any node $w\in \bar{R}$, we can determine its parent and children in $\Vor(w)$ at query and time complexity $\tilde{O}(n^{1/3})$.
We conclude that we can determine $p(w)$ and whether $T(w)<k$ or $T(w)\geq k$ at query complexity $\tilde{O}(n^{2/3})$, by partially or completely exploring $T(w)$;
if $T(w)<k$, we determine $T(w)$ completely.
We collect this information for $u$ and its $\ell-1$ ancestors, and determine the cluster $Q$ of $u$.
Finally, we repeat the procedure to reconstruct the cluster $W$ of $v$, and determine for all nodes adjacent to either cluster whether they are in $\bar{R}$ and, if so, their centers.
The query and time complexity of this operation is $\tilde{O}(n^{2/3})$ in total.

For the cases $A=Q$, $B=W$ and $A=W$, $B=Q$, respectively, with the above information we can determine
\begin{itemize}
  \item whether $\Vor(u)=\Vor(v)$,
  \item if $\Vor(u)=\Vor(v)$, whether $p(u)=v$ or $p(v)=u$,
  \item if $\Vor(u)\neq \Vor(v)$,
  \begin{itemize}
    \item whether $A$ is marked (i.e., $c(A)$ has been marked) and whether $\{u,v\}$ has minimum rank in $E(A,B)$,
    \item whether $A$ is not adjacent to any marked cluster (i.e., none of the adjacent nodes' centers has been marked) and whether $\{u,v\}$ has minimum rank in $E(A,\Vor(B))$, and
    \item whether there is a marked cluster $C$ adjacent to $A$ so that $A$ participates in $\calC(C)$, $\Vor(A)$ has minimum rank in $\Vor(\partial B)\cap \Vor(\partial C)$, and $\{u,v\}$ has minimum rank in $E(B,\Vor(A))$. We note that since we have degree bounded by $\Delta$, the number of vertices in $A$ is $\tilde{O}(n^{1/3})$, and the probability that a cell is marked is $n^{-1/3}$, the number of cluster-of-clusters that $A$ participates in is $\tilde{O}(\Delta)$ w.h.p.  
  \end{itemize}
\end{itemize}
In other words, we can perform all necessary checks to decide whether $\{u,v\}\in E'$ or not.

The algorithm in Section~\ref{sec:promise} requires $\tilde{O}(n^{2/3})$ random bits for the selection of centers and marked clusters. 
For the emulation of the algorithm of Elkin and Neiman it suffices that the random variables $\{r_u\}$ be $\tilde{O}(n^{1/3})$-wise independent, because the outcome of the algorithm for a vertex $v \in R$ depends only on the random variables of at most $\tilde{O}(n^{1/3})$ vertices (the vertices in $\Gamma_\ell(v)$).
Thus, overall $\tilde{O}(n^{2/3})$ random bits (up to $\poly(\Delta/\eps)$ factors) are sufficient. 
\EPF

\bibliographystyle{plain}
\bibliography{refs2}

\end{document}

%% file: abstract.tex
Constructing a sparse \emph{spanning subgraph} is a fundamental primitive in graph theory.
In this paper, we study this problem in the Centralized Local model, where the goal is to decide whether an edge is part of the spanning subgraph by examining only a small part of the input;
yet, answers must be globally consistent and independent of prior queries.

Unfortunately, maximally sparse spanning subgraphs, i.e., spanning trees, cannot be constructed efficiently in this model.
Therefore, we settle for a spanning subgraph containing at most $(1+\epsilon)n$ edges (where $n$ is the number of vertices and $\epsilon$ is a given approximation/sparsity parameter).
We achieve a query complexity of $\tilde{O}(\poly(\Delta/\varepsilon)n^{2/3})$,\footnote{$\tilde{O}$-notation hides polylogarithmic factors in $n$.} where $\Delta$ is the maximum degree of the input graph.
Our algorithm is the first to do so on arbitrary bounded degree graphs.
Moreover, we achieve the additional property that our algorithm outputs a \emph{spanner,} i.e., distances are approximately preserved.
With high probability, for each deleted edge there is a path of $O(\log n\cdot (\Delta+\log n)/\epsilon)$ hops in the output that connects its endpoints.

%% file: intro.tex
\section{Introduction}

When operating on very large graphs, it is often impractical or infeasible to (i) hold the entire graph in the local memory of a processing unit, (ii) run linear-time (or even slower) algorithms, or even (iii) have only a single processing unit perform computations sequentially.
These constraints inspired the Centralized Local model~\cite{RTVX}, which essentially views the input as being stored in a (likely distributed) database that provides query access to external processing units.
To minimize the coordination overhead of such a system, it is furthermore required that there is no shared memory or communication between the querying processes, except for shared randomness provided alongside the access to the input.
The idea is now to run sublinear-time algorithms that extract useful global properties of the graph and/or to examine the input graph locally upon demand by applications.

Studying graphs in this model leads to the need for query access to a variety of graph-theoretical structures like, e.g., independent or dominating sets.
In such a case, it is crucial that locally evaluating whether a node participates in such a set is \emph{consistent} with the same evaluation for other nodes.
This is a non-trivial task, as local decisions can only be coordinated implicitly via the structure of the input (which is to be examined as little as possible) and the shared randomness.
Nonetheless, this budding field brought forth a number of elegant algorithms solving, e.g., maximal independent set, hypergraph coloring, $k$-CNF, approximate maximum matching and approximate minimum vertex cover for bipartite graphs~\cite{ARVX12,EMR14,FMS15,LRY16,MRVX12,MV13,RTVX}.

In this work, we consider another very basic graph structure: sparse spanning subgraphs.
Here, the task is to select a subset of the edges of the (connected) input graph so that the output is still connected, but has only few edges.
By ``few'' we mean that, for some input parameter $\varepsilon>0$, the number of selected edges is at most $(1+\varepsilon)n$, where $n$ denotes the number of nodes.
One may see this as a relaxed version of the problem of outputting a spanning tree of the graph, which is a too rigid requirement when looking for fast algorithms:
on a cycle, a single edge has to be deleted, but this necessitates to first verify that the input graph is not, in fact, a line.

\BD[\cite{LRR14}] 
\label{dfn:SSG-alg}
An algorithm $\calA$ is a {\em Local Sparse Spanning Graph (LSSG) algorithm} if, given $n,\Delta \ge 1$, a parameter $\eps \geq 0$, and query access to the incidence list representation of a connected graph $G=(V,E)$ over $n$ vertices and of degree at most $\Delta$, it provides oracle access to a subgraph $G'=(V, E')$ of $G$ such that:
\BE
\item\label{it:connect} $G'$ is connected. 
\item\label{it:internal-rand} $\s{E'} \leq (1+\eps)\cdot n$ with high probability (w.h.p.),\footnote{That is, with probability at least $1-1/n^c$ for an arbitrary constant $c>0$ that is chosen upfront.}
where $E'$ is determined by $G$ and the internal randomness of~$\mathcal{A}$.
\EE
By ``providing oracle access to $G'$'' we mean that on input $\{u,v\}\in E$, $\calA$ returns whether $\{u,v\} \in E'$, and for any sequence of edges, $\calA$ answers consistently with respect to the same~$G'$.
\ED

We are interested in LLSG algorithms that, for each given edge, perform as few queries as possible to $G$.
Observe that Item~\ref{it:internal-rand} implies that the answers of an LLSG algorithm
to queries cannot depend on previously asked queries.

We note that relaxing from requiring a tree as output makes it possible to ask for additional guarantees that, in general, cannot be met by a spanning tree.
Instead of merely preserving connectivity, it becomes possible to maintain \emph{distances} up to small factors.
Such subgraphs are known as (sparse, multiplicative) \emph{spanners}~\cite{PS89,PU89}.
In fact, choosing $\varepsilon \in o(1)$ then yields \emph{ultra-sparse} spanners that are $o(n)$ edges away from being trees.

\subsection{Our Contribution}

We give the first non-trivial LSSG algorithm in the Centralized Local model that runs on arbitrary graphs.
We achieve a query complexity of $\tilde{O}(\poly(\Delta/\varepsilon)n^{2/3})$ per edge, w.h.p.
Moreover, we guarantee that for each edge that is not selected into the spanner, w.h.p.\ there is a path of $O(\log n\cdot (\Delta+\log n)/\eps)$ hops consisting of edges that are selected into the spanner;
this is referred to as a \emph{stretch} of $O(\log n\cdot (\Delta+\log n)/\eps)$.

For simplicity, assume for the moment that $\Delta$ and $\varepsilon$ are constants. Our algorithm combines the following key ideas.
\begin{compactitem}
  \item We classify edges as expanding if there are sufficiently many (roughly $n^{1/3}$) nodes within $O(\log n)$ hops of their endpoints.
  For non-expanding edges, we can efficiently simulate a standard distributed spanner algorithm at small query complexity, as solutions of running time $O(\log n)$ are known~(e.g.~\cite{EN17}).
  \item On the node set induced by expanding edges, we can construct a partition into Voronoi cells with respect to roughly $n^{2/3}$ randomly sampled centers.
  The Voronoi cells are spanned by trees of depth $O(\log n)$, as expanding nodes have their closest center within $O(\log n)$ hops w.h.p.
  Finding the closest center has query complexity $\tilde{O}(n^{1/3})$.
  \item We refine the partition into Voronoi cells further into \emph{clusters} of $\tilde{O}(n^{1/3})$ nodes.
  We simply let a node be a singleton cluster if its subtree in the spanning tree of its cell contains more than $\tilde{O}(n^{1/3})$ nodes.
  This construction has query complexity $\tilde{O}(n^{2/3})$ for constructing a complete cluster, yet ensures that there are $\tilde{O}(n^{2/3})$ clusters in total due to the low depth of the trees;
  moreover, each cluster is completely contained in some Voronoi cell.
  \item It remains to select few edges to interconnect the Voronoi cells.
  This is the main challenge, for which the above properties of the partition are crucial.
  To keep the number of selected edges small in expectation, we mark a subset of expected size $\tilde{\Theta}(n^{1/3})$ of the clusters by marking each Voronoi cell (and thereby its constituent clusters) with probability $n^{-1/3}$.
  We then try to ensure that (i) clusters select an edge to each adjacent marked Voronoi cell and (ii) for each marked Voronoi cell adjacent to an adjacent cluster, they select one edge connecting to \emph{some} cluster adjacent to this cell.
  \item The main issue with the previous step is that we cannot afford to construct each adjacent cluster, preventing us from guaranteeing (ii).
  We circumvent this obstacle by identifying for adjacent clusters in which cell they are and keeping an edge for the purpose of (ii) if it satisfies a certain minimality requirement with respect to the \emph{rank} of the cell used for symmetry breaking purposes.
  This way, we avoid construction of adjacent clusters, instead needing to determine the rank of their Voronoi cells only.
  This way, we maintain query complexity $\tilde{O}(n^{2/3})$.
  \item However, this now entails an inductive argument for ensuring connectivity, which also affects stretch.
  By choosing Voronoi cell ranks uniformly at random, we ensure that the length of such an inductive chain is bounded by $O(\log n)$ w.h.p. 
  Together with the depth of the Voronoi cell trees of $O(\log n)$ and the stretch of the spanner algorithm for non-expanding edges (also $O(\log n)$), this yields the total bound of $O(\log^2 n)$ on the stretch of our scheme.
\end{compactitem}

Finally, we note that we can place the above routine in a wrapper verifying that, w.h.p., the number of globally selected edges does not significantly exceed the expectation.
If this is not the case, the wrapper starts the process all over.
Since in each attempt the success probability is constant (and the verifier succeeds w.h.p.), we get within $O(\log n)$ attempts that the bound on the number of selected edges is satisfied w.h.p.\ and the routine terminates.

\subparagraph*{Relation to Property Testing.}
As observed in~\cite{LRR14-full}, testing cycle-freeness with one sided-error in the bounded-degree model can be reduced to the LSSG problem.
From this reduction it follows that we obtain a tester for cycle-freeness that works in $\tilde{O}(n^{2/3})$ time.
Czumaj et al.~\cite{CGRSSS} studied the problem of $C_k$-minor freeness with one sided-error. For cycle-freeness ($C_3$-minor freeness) their complexity is $\tilde{O}(\sqrt{n})$, therefore their complexity is better than ours.
However, we would like to point out that an LSSG algorithm gives a stronger guarantee than a one-sided error tester for cycle-freeness.
An LSSG algorithm can be used to find for all but at most $(1+\eps)|V|$ edges $e$ a cycle that $e$ belongs to, i.e., a witness for the violation by $e$ can be provided.
In constrast, a one-sided error tester merely guarantees to find a single cycle in instances that are $\eps$-far from being cycle-free.  

Perhaps more importantly, our approach proves useful when testing for other minors.
Recently, Fichtenberger et al.~\cite{FLVW17} built on our work to construct one-sided error testers for each minor of the $(k \times 2)$-grid.
\subsection{Related work}
\input{related-work}

%% file: related-work.tex
The problem of finding a sparse spanning subgraph in the Centralized Local model was first studied in~\cite{LRR14}, where the authors show a lower bound of $\Omega(\sqrt{n})$ queries for constant $\eps$ and $\Delta$ (see also survey by Rubinfeld~\cite{Rub17}).
They also present an upper bound with nearly tight query complexity for graphs that have very good expansion properties. 
However, for general (bounded degree) graphs their algorithm might query the entire graph for completing a single call to the oracle.
They also provide an efficient algorithm for minor-free graphs that was later improved in~\cite{LR15}. 
The algorithm presented in~\cite{LR15} achieves a query complexity that is polynomial in $\Delta$ and $1/\eps$ and is independent of $n$.
The stretch factor of this algorithm is also independent of $n$ and depends only on $\Delta$, $1/\eps$, and the size of the excluded minor.  

A characterization of the query complexity of the problem was presented in~\cite{LMRRS}.
Specifically, this work provides an upper bound (which builds on an algorithm in~\cite{LRR14}) that has a query complexity that is independent of $n$ (however, super-exponential in $1/\eps$) for families of graphs which are, roughly speaking, sufficiently non-expanding everywhere.
On the other hand, they show that, for a family of graphs with expansion properties that are slightly better, any local algorithm must have a query complexity that depends on $n$.

In the (distributed) Local model, Ram and Vicari~\cite{Rm2011} study the same problem and provide an algorithm that runs in $\min\{D, O(\log n)\}$ rounds, where $D$ denotes the diameter of the input graph. Their algorithm achieves the sparsity property by breaking all short cycles. 

%% file: conn.tex
\subsection{Connectivity and Stretch} 
\BL\label{lem:adj}
$H$ is connected.
\EL
\BPF
Recall that $H$ contains a spanning tree on every Voronoi cell, hence it suffices to show that we can connect any pair of Voronoi cells by a path between some of their vertices.
Moreover, the facts that $G$ is connected and the Voronoi cells are a partition of $V$ imply that it is sufficient to prove this for any pair of adjacent Voronoi cells.
Accordingly, let $\Vor$ and $\Vor_1$ be two cells such that $E(\Vor, \Vor_1) \neq \emptyset$.

Consider clusters $A\subseteq \Vor$ and $B\subseteq \Vor_1$ such that the edge $e$ of minimum rank in $E(\Vor,\Vor_1)$ is in $E(A,B)$.
If $B$ is not adjacent to a marked cell, then Condition~\ref{connect_no_adjacent} implies that $e$ is selected into $H$.
Thus, we may assume that $B$ is adjacent to a marked cell $\Vor'$.
Accordingly, there exists a marked cluster $C\subseteq \Vor'$ such that $B$ is participating in $\mathcal{C}(C)$.

If the rank of $\Vor_1$ is minimum in $\Vor(\partial C)\cap \Vor(\partial A)$, then $e$ is selected into $H$ by Condition~\ref{connect_indirect} and we are done.
Otherwise, observe that $\Vor_1$ is connected to $\Vor'$, as the edge of minimum rank in $E(B,C)$ is selected into $H$ by Condition~\ref{connect_marked}.
Therefore, it suffices to show that $\Vor$ gets connected to $\Vor'$.
Let $\Vor_2$ be the cell of minimum rank among $\Vor(\partial C)\cap \Vor(\partial A)$.
Let $D\subseteq \Vor_2$ be the cluster satisfying that the edge $e'$ of minimum rank in $E(A,\Vor_2)$ is in $E(A,D)$.
Note that $\Vor_2$ is connected to $\Vor'$ (which we saw to be connected to $\Vor_1$), as there is some cluster $D'\subseteq \Vor(D)$ that is adjacent to $C$ and selects the edge of minimum rank in $E(D',C)$ by Condition~\ref{connect_marked}.

Overall, we see that it is sufficient to show that $\Vor$ gets connected to $\Vor_2$, where $\Vor_2$ has smaller rank than $\Vor_1$.
We now repeat the above reasoning inductively.
In step $i$, we either succeed in establishing connectivity between $\Vor$ and $\Vor_i$, or we determine a cell $\Vor_{i+1}$ that has smaller rank than $\Vor_i$ and is connected to $\Vor_i$.
As any sequence of Voronoi cells of descending ranks must be finite, the induction halts after finitely many steps.
Because the induction invariant is that $\Vor_{i+1}$ is connected to $\Vor_i$, this establishes connectivity between $\Vor$ and $\Vor_1$, completing the proof.
\EPF

\BL\label{lem:chain}
Denote by $G_{\Vor}$ the graph obtained from $G$ by contracting Voronoi cells and by $H_{\Vor}$ its subgraph obtained when doing the same in $H$.
If the cells' ranks are uniformly random, w.h.p.\ $H_{\Vor}$ is a spanner of $G_{\Vor}$ of stretch $O(\log n)$.
\EL
\BPF
Recall the proof of Lemma~\ref{lem:adj}.
We established connectivity by an inductive argument, where each step increased the number of traversed Voronoi cells by two.
Hence, it suffices to show that the induction halts after $O(\log n)$ steps w.h.p.

To see this, observe first that $G_{\Vor}$ is independent of the ranks assigned to Voronoi cells and pick any pair of adjacent cells $\Vor$, $\Vor_1$, i.e., neighbors in $G_{\Vor}$.
We perform the induction again, assigning ranks from high to low only as needed in each step, according to the following process.
In each step, we query the rank of some cells, and given an answer of rank $r$, the ranks of all cells of rank at least $r$ are revealed as well.  
In step~$i$, we begin by querying the rank of $\Vor_i$.
Consider the cluster $D_i \subseteq \Vor_i$ adjacent to $A$ satisfying that the edge with minimum rank in $E(\Vor_i, A)$ is also in $E(D_i, A)$.
We can assume without loss of generality that $D_i$ is adjacent to a marked cluster $F_i$ and that it is participating in $\mathcal{C}(F_i)$ (as otherwise $D_i$ connects to $A$ directly and we can terminate the process). 
If the ranks of all the cells adjacent to both $F_i$ and $A$ were already revealed, then the process terminates. 
Otherwise, we query the rank of all such cells whose rank is still unrevealed.
We set the cell of the queried cluster that has minimum rank to be $\Vor_{i+1}$ and we continue to the next step.

We claim that, in each step $i$, either the process terminates, or the rank of $\Vor_{i+1}$ is at most half of the rank of $\Vor_i$ with probability at least $1/2$.
To verify this, observe that in the beginning of step $i$, any cell center whose rank was not revealed so far has rank which is uniformly distributed in $[r_i-1]$, where $r_i$ is the rank of $\Vor_i$.\footnote{In step $1$, we first query $\Vor_1$ and then observe that this statement holds.}
With probability at least $1/2$, such a rank is at most $r_i/2$.
If $\Vor_i$ has no adjacent cells whose ranks have not been revealed yet, the process terminates.
Hence, regardless of whether the process terminates or not, the claim holds.

By Chernoff's bound, we conclude that the process terminates within $O(\log n)$ steps w.h.p., as $r_1$ is bounded by the number of Voronoi cells, which itself is trivially bounded by $n$.
By the union bound over all pairs of cells $\Vor$ and $\Vor_1$, we get the desired guarantee. 
%
\EPF

\BC\label{cor:chain}
W.h.p., $H$ is a spanner of $G$ of stretch $O(\log n\cdot (\Delta+\log n)/\eps)$.
\EC
\BPF
Due to the promise on $G$, w.h.p.\ the spanning trees on Voronoi cells have depth $\ell\in O((\Delta+\log n)/\eps)$.
Hence, for any edge within a Voronoi cell, the claim holds w.h.p.
Moreover, for an edge connecting different Voronoi cells, by Lemma~\ref{lem:chain}, w.h.p.\ there is a path of length $O(\log n)$ in $H_{\Vor}$ connecting the respective cells.
Navigating with at most $2\ell\in O((\Delta+\log n)/\eps)$ hops in each traversed cell, we obtain a suitable path of length $O(\log n\cdot (\Delta+\log n)/\eps)$ in $H$. 
\EPF

%% file: alg-general.tex
\section{The Algorithm for General Graphs}
We use a combination of the algorithm in Section~\ref{sec:promise} with the algorithm by Elkin and Neiman for ultra-sparse spanners~\cite{EN17}.
We call a vertex $v$ {\em remote}, with respect to a set of centers, if the $\ell$-hop neighborhood of $v$ does not include a center. 
Fix $S$, let $R$ denote the set of remote vertices with respect to $S$, and abbreviate $\bar{R}\eqdef V\setminus R$.

\subparagraph*{First Step.} Run the algorithm from Section~\ref{sec:promise} on the subgraph induced by $\bar{R}$, i.e., $\{u,v\}\in E$ with $u,v\in \bar{R}$ is added to $E'$ if and only if the algorithm outputs the edge.

\subparagraph*{Second Step.} Run the algorithm of Elkin and Neiman~\cite{EN17} on the subgraph induced by $R$, i.e., $\{u,v\}\in E$ with $u,v\in R$ is added to $E'$ if and only if the algorithm outputs the edge.\footnote{The algorithm is described for connected graphs; we simply apply it to each connected component of $R$.}
Their algorithm proceeds as follows. Given an integer $h$, each vertex $v$ draws $r_v$ according to the exponential distribution with parameter $\beta = \ln (n/\delta)/h$, where $\delta$ is a parameter that controls the success probability of the algorithm.
Each vertex $v$ receives $r_u$ from every vertex $u$ within distance $h$, and stores $m_u(v) = r_u - \d(u ,v)$ and a neighbor on a shortest path between $v$ and $u$, denoted by $n_u(v)$.
The edges that are added to the spanner are $C(v) = \{\{v, n_u(v)\}\,|\, m_u(v) \geq \max_{w\in V}\{m_w(v) -1\}\}$, for every $v\in R$.
With probability at least $1-\delta$, it holds that $r_v < h$ for all $v \in V$ (Claim 2.3 in~\cite{EN17}). 
We choose $\delta = 1/n^c$ (i.e., the algorithm succeeds w.h.p.) and $h = \ell$, where w.l.o.g.\ we require $c\geq 1$.
The following lemma implies that the total number of edges that we add to $H$ in the second step is at most $|R| \cdot (n^2)^{1/\ell} \leq |R| (1+\eps)$ in expectation.
\BL[Proof of Lemma 2.2 in~\cite{EN17}]\label{lem:stretch_EN}
For every $v\in R$, $\Exp[C(v)] \leq (n/\delta)^{1/h}$.
\EL

\subparagraph*{Third Step.} Add to $E'$ all edges $e\in E(R,\bar{R})$. 

\medskip\noindent\\
The following lemma implies that the expected number of edges which are added in the third step is at most $\eps n$.

\BL
$\Exp[|E(R,\bar{R})|]\leq \eps n$.
\EL
\BPF
Observe that for an edge $\{u,v\}\in E$, there is at most one integer $r$ such that $\Gamma_{r}(u) \cap S = \emptyset$ and $\Gamma_{r}(v) \cap S \neq \emptyset$ (or vice versa).
If there is no such $r$ or $\ell \neq r$, then the edge is not in $E(R,\bar{R})$. Over the random choice of $\ell$, the probability of the event that the edge is included is at most $\Pr[\ell = r] \leq \eps/\Delta$. The desired bound follows by linearity of expectation.
\EPF

\subsection{Stretch Factor}
Consider any edge $e=\{u,v\}\in E\setminus E'$ we removed.
If $u,v\in \bar{R}$, $e$ was removed by the Algorithm from Section~\ref{sec:promise}, which was applied to the subgraph induced by $R$.
Applying Lemma~\ref{lem:chain} to the connected component of $e$, we get that w.h.p.\ there is a path of length $O(\log n\cdot (\Delta+\log n)/\eps)$ from $u$ to $v$ in $H$.
If $u,v\in R$, by Lemma~\ref{lem:stretch_EN} and the choice of parameters, w.h.p.\ there is a path of length $O(\log n)$ from $u$ to $v$ in $H$.
As $e\notin E(R,\bar{R})$ by the third step, we arrive at the following corollary.
\begin{cor}
The above algorithm guarantees stretch $O(\log n\cdot (\Delta+\log n)/\eps)$ w.h.p.\ and satisfies that $\Exp[|E'|]\in (1+O(\eps))n$.
\end{cor}